\documentclass[12pt, a4paper]{JHEP}

\usepackage{amssymb}
\usepackage{epsfig}
\usepackage{psfrag}

\newcommand{\Tr}{\mathop{\rm Tr}\nolimits}

\def\b{\beta}
\def\d{\delta}
\def\e{\epsilon}
\def\f{\phi}
\def\g{\gamma}
\def\h{\eta}
\def\p{\pi}
\def\s{\sigma}

\def\x{\xi}
\def\F{\Phi}

\def\ch{{\cal H}}

\def\be{\begin{equation}}
\def\ee{\end{equation}}
\def\bea{\begin{eqnarray}}
\def\eea{\end{eqnarray}}
\def\pa{\partial}

\newcommand{\hp}{{\widehat\Phi}}
\newcommand{\hq}{{\widehat Q_B}}
\newcommand{\ha}{{\widehat{A}}}
\newcommand{\llll}{\left\langle\left\langle\! \right.\right.}
\newcommand{\rrr}{\left.\left.\!\right\rangle\right\rangle}
\newcommand{\STRUT}{\rule{0in}{4ex}}

\title{The tachyon potential in Witten's superstring field theory}

\author{Pieter-Jan De Smet and Joris Raeymaekers
\\Instituut voor theoretische fysica, Katholieke
Universiteit Leuven\\
Celestijnenlaan 200D, B-3001 Leuven, Belgium \\
E-mail: \email{Joris.Raeymaekers@fys.kuleuven.ac.be},
\email{Pieter-Jan.DeSmet@fys.kuleuven.ac.be} }

\abstract{We study the tachyon potential in the NS sector of
Witten's cubic superstring field theory. In this theory, the pure
tachyon contribution to the potential has no minimum. We find
that this remains the case  when higher modes up to level two are
included.}

\keywords{D-branes, Superstring Vacua}
\preprint{ KUL-TF-2000/14 \\
  {\tt hep-th/0004112}}
\received{May 22, 2000}

\accepted{August 18, 2000\smash{\llap{\raise-1\baselineskip\hbox{{\sc
Revised:} July 10, 2000}}}}

\begin{document}

\section{Introduction}

It has been conjectured by Sen that at the stationary point of
the tachyon potential for the D-brane-anti-D-brane pair or for
the non-BPS D-brane of superstring theories, the negative energy
density  precisely cancels the brane tensions~\cite{Sen}.

For the D-brane of bosonic string theory, this conjecture has
been verified~\cite{bosonic} starting from Witten's open string
field theory~\cite{WITTENBSFT} and in the supersymmetric case
starting from Berkovits' superstring field
theory~\cite{actieBerko}--\cite{Iqbal}.\footnote{We do not agree
with the results of~\cite{Iqbal}.}

In this paper we study  the tachyon potential in the NS sector of
Witten's superstring field theory~\cite{WITTENSFT}. Soon after
this theory was proposed, it became clear that it suffered some
problems~\cite{wendt}. Infinities were found to arise in the
calculation of tree-level scattering amplitudes. These find their
origin in picture-changing operators inserted at the same point.
Similar problems arise in the proof of gauge invariance of the
action. Modifications of Witten's action have been proposed in
order to solve these issues~\cite{preitschopf}, but these seem to
suffer from other difficulties~\cite{actieBerko}. In the light of
these problems, the present calculation should be seen as a
further study of Witten's string field theory  proposal rather
than as a test of Sen's conjecture, which has been extensively
verified on a quantitative level.

We use the level truncation method of Kostelecky and
Samuel~\cite{KS} retaining fields up to level~2 and terms up to
level~4 in the action. We find that the tachyon potential in this
theory does not have the behaviour conjectured by Sen, in fact it
doesn't even have a minimum. This is already the case for the
pure tachyon contribution (in contrast to the pure tachyon
contribution is Berkovits' theory~\cite{0001084}), and this
behaviour is found to persist  when higher modes are included.

\section{Witten's superstring field theory in CFT language}

In this section we review Wittens string field theory for open
superstrings~\cite{WITTENBSFT,WITTENSFT} formulated in  the
conformal field theory (CFT) approach~\cite{leclair}. We restrict
our attention to fields in the NS sector of the theory.

In the CFT approach, a general NS-sector string field $A$ is
represented by a CFT vertex operator of ghost number $1$ and
picture number $-1$. The superghosts are bosonized as $\b =
\partial  \x e^{-\f}$ and $\g =
\h e^\f$~\cite{FMS} and the string field $A$ is restricted to
live in the ``small'' Hilbert space of the bosonized system, i.e.\
it does not include the $\x$ zero mode. The string field should
also satisfy a suitable reality condition in order for the action
to take real values.

Witten proposed a Chern-Simons type action
\be
S = -\llll A \  Q_BA\rrr - \frac{2 }{ 3}\llll A^3 \rrr,
\label{Waction}
\ee
where $Q_B$ is the BRST charge
$$
Q_B = \oint dz j_B(z)
= \oint dz \Bigl\{  c \bigl( T_m + T_{\xi\eta} + T_\phi)
+ c \partial c b +\eta \,e^\phi
\, G_m - \eta\partial \eta e^{2\phi} b \Bigr\}\, ,
$$
and the double brackets should be interpreted as a CFT
correlator\footnote{The elementary CFT correlator is normalized
as $\langle c \pa c  \pa^2 c (z_1) e^{-2 \f}(z_2) \rangle =2 $ }
with extra picture-changing insertions:
\begin{eqnarray}
\label{dbrackets}
& &\llll \F_1 \F_2 \cdots \F_{n-1} \F_n \rrr =\\
& &=
\Big\langle Y(0)
f_1^n \circ \F_1(0) X(0)
f_2^n \circ \F_2(0) X(0)  \cdots X(0) f_{n-1}^n \circ
\F_{n-1}(0)X(0) f_{n}^n
\circ \F_{n}(0) \Big\rangle\,.\nonumber
\end{eqnarray}
The symbols $f_i^n$ denote conformal transformations mapping the
unit circle to wedge-formed pieces of the complex-plane: $$
f^{n}_k(z) = e^{2\pi i (k-1)/ n} \left(\frac{1+iz}{ 1-iz}
\right)^{2/n} \quad  \hbox{for} \ n\geq 1\,.
$$ The operators $X$ and $Y$ are the picture-changing  and inverse
picture-changing operators, respectively:
\begin{eqnarray}
X &=& - \pa \x c + e ^\f G_m - \pa \h\  b   e^{2\f}- \pa\left(\h  b
 e^{2\f}\right),\nonumber\\
Y &=& - \pa \x c e^{- 2 \f}\,.\nonumber
\end{eqnarray}
They are each others inverse in the following sense:
$$
\lim_{\e \rightarrow 0} Y(z + \e) X(z) =1\,.
$$
The action is formally  invariant under the
gauge transformations
$$
\d A = Q_B \e + A \e - \e A\,,
$$
where $\e$ is a NS string field of ghost number 0 in the $-1$
picture.
The proof of the gauge invariance relies on the following
properties of the
correlator~(\ref{dbrackets})
\bea
\llll Q_B (\F_1 \cdots \F_n) \rrr &=& 0 \,, \nonumber\\
\llll \cdots Q_B^2  (\F_1 \cdots \F_n) \cdots \rrr &=& 0
\,, \nonumber\\
\llll \cdots Q_B (A \F) \cdots \rrr &=& \llll \cdots( Q_B A\
\F - A\ Q_B \F) \cdots \rrr, \nonumber\\
\llll \cdots Q_B (\e \F) \cdots \rrr &=& \llll \cdots (Q_B \e\
\F + \e\ Q_B \F) \cdots \rrr, \nonumber\\
\llll \F_1 \cdots \F_{n-1} \F_n \rrr &=& \llll \F_n \F_1 \cdots
 \F_{n-1} \rrr. \label{algstruct}
\eea
$A$
and $\e$ represent the string field and the gauge parameter
respectively and the $\F_i$ represent arbitrary operators in the
GSO(+) sector.

These properties suffice to prove the gauge invariance of the
action. Such a ``proof'' should be taken with a grain of salt
however since it involves manipulating correlators with four
string fields. From the definition~(\ref{dbrackets}), one sees
that such correlators contain 2 insertions of the
picture-changing operator at the origin and are hence
divergent.\footnote{A similar problem is encountered in the proof
of associativity of Witten's star product of string~fields.}
These and other problems of the action~(\ref{Waction}) were
discussed in~\cite{wendt}.

\section{Witten's string field theory on a non-BPS D-brane}

In order to describe NS excitations on a non-BPS D-brane, one
should extend the theory to include also states with odd
world-sheet fermion number. This can be accomplished by tensoring
the fields entering in the action and gauge transformation  with
suitable internal Chan-Paton (CP) factors as
in~\cite{0001084,BSZ}. The fields with CP factors attached are
denoted with hats. We take the following assignments for the
internal CP factors:
\begin{eqnarray}
\ha &=& A_+ \otimes I + A_- \otimes \s_1\,,\nonumber\\
\widehat{\e} &=& \e_+ \otimes \s_3 + \e_- \otimes  i \s_2
\,,\nonumber\\
\nonumber
\hq &=& Q_B \otimes \s_3\,, \\
\widehat{Y} &=& Y \otimes I\,, \nonumber\\
\widehat{X} &=& X \otimes \s_3\,,\label{CPfactors}
\end{eqnarray}
where $I$ is the $2\times 2$ unit matrix and $\s_i$ are the Pauli
matrices. The action takes the~form
$$
S = -\llll \ha \
\hq\ha\rrr - \frac{2 }{ 3} \llll\ha^3 \rrr,
$$
where the double
brackets should now be interpreted as
\begin{eqnarray}
\llll \hp_1 \hp_2 \cdots \hp_{n-1} \hp_n \rrr& =&
\frac{1 }{ 2} \Tr \Big\langle
\widehat{Y}(0)
f_1^n \circ \hp_1(0)\widehat{X}(0) f_2^n \circ
\hp_2(0)\times\nonumber\\
&&\hphantom{\frac{1 }{ 2} \Tr \Big\langle}\!\times\,\widehat{X}(0)  \cdots \widehat{X}(0)
f_{n-1}^n \circ
\hp_{n-1}(0) \widehat{X}(0) f_{n}^n
\circ \hp_{n}(0) \Big\rangle\,.
\nonumber
\label{dbrackets2}
\end{eqnarray}
The trace runs over the internal CP indices. The gauge
transformations are
\be
\d \ha = \hq \widehat{\e} + \ha \widehat{\e} - \widehat{\e} \ha\,.
\label{gaugeinv}
\ee
The CP factor assignments~(\ref{CPfactors})
were chosen such  that gauge transformations
preserve the CP structure of the string field:
$$
\d \ha = \d A_+ \otimes I + \d A_- \otimes \s_1
$$
and such that the algebraic structure~(\ref{algstruct}) is preserved:
\begin{eqnarray}
\llll \hq (\hp_1 \cdots \hp_n) \rrr &=& 0\,, \nonumber\\
\llll \cdots \hq^2  (\hp_1 \cdots \hp_n) \cdots \rrr &=& 0\,,
\nonumber\\
\llll \cdots \hq (\ha \hp) \cdots \rrr &=& \llll \cdots( \hq \ha\
\hp - \ha\ \hq \hp) \cdots \rrr ,\nonumber\\
\llll \cdots \hq (\widehat{\e} \hp) \cdots \rrr &=& \llll
\cdots (\hq
\widehat{\e}\
\hp + \widehat{\e}\ \hq \hp) \cdots \rrr ,\nonumber\\
\llll \hp_1 \cdots \hp_{n-1} \hp_n \rrr &=& \llll \hp_n \hp_1 \cdots
 \hp_{n-1} \rrr. \nonumber
\end{eqnarray}
With these properties (proven in appendix~\ref{proofcyclicity}),
the (formal) proof of gauge invariance goes through as in the
GSO(+) sector.

\section{The fields up to level 2}

In the calculation of the tachyon potential, we can restrict the
string field to lie in a subspace $ \ch_1$ formed by acting only
with modes of the stress-energy tensor, the supercurrent and the
ghost fields $b,\ c,\ \eta,\ \x, \f $ since the other excitations
can be consistently put to zero~\cite{Sen}. We fix the gauge
freedom~(\ref{gaugeinv}) by imposing the Feynman-Siegel gauge $
b_0 \ha  = 0 $ on  fields with non-zero conformal weight. Taking
all this together we get the following list of contributing
fields up to level 2. The level of a field is just the conformal
weight shifted by 1/2, in this way the tachyon is a  level 0
field. We use the notation $|q\rangle$ for the state corresponding
with the operator $e^{q \phi}$.

\begin{table}[h]
\caption{The level $0$,$1$ and $2$ fields should be tensored
with $\sigma_1$ and the level $1/2$ and $3/2$ fields with~$I$. We
list the conformal transformations of the fields in
appendix~\ref{transforms}.}
\begin{center}
\begin{tabular}{|l|c|c|l|}\hline
\mbox {Level}
& \mbox {GSO}&\mbox{state}& \mbox {vertex\ operator} \\
\hline
0& $-$ & $ c_1 |-1 \rangle $ & $ T = c e^{-\phi}$\\
\hline
1/2\STRUT& $+$ & $\xi_{-1} c_1 c_0 |-2\rangle$&
$ R = \partial\xi c\partial c \ e^{-2\phi}$\\
\hline
1\STRUT& $-$ & $ c_1 \phi_{-1}|-1\rangle $  &
$ S =  c\partial\phi\   e^{-\phi}$\\
\hline
3/2\STRUT& $+$ &$ 2 c_1 c_{-1} \xi_{-1} |-2\rangle$
&$A = c\partial^2 c\partial\xi\ e^{-2\phi}$\\
& &$ \eta_{-1}  |0\rangle$&$E =  \eta$\\
& &$  c_1 G^m_{-3/2}  |-1\rangle$&$F = c G^m\ e^{-\phi}$\\
\hline
2\STRUT& $-$ &$ c_1 \left[(\phi_{-1})^2-\phi_{-2}\right]|-1\rangle$
& $ K =  c \ \partial^2\left(e^{-\phi}\right)$\\
& &$  c_1 \phi_{-2} |-1\rangle$&$ L =  c\ \partial^2\phi\
e^{-\phi}$\\
& &$c_1 L^m_{-2}  |-1\rangle$&$ M = c T^m\ e^{-\phi}$\\
& & $2  c_{-1}   |-1\rangle$& $N =  \partial^2 c\ e^{-\phi}$\\
& &$ \xi_{-1}\eta_{-1}c_1  |-1\rangle$&$ P = \partial\xi\eta c\
e^{-\phi}$\\\hline
\end{tabular}
\end{center}
\end{table}

\section{The tachyon potential}

We have calculated the tachyon potential involving fields up to
level 2, including only the terms up to level 4 (the level of a
term in the potential is defined to be the sum of the levels of
the fields entering into it). We have performed the actual
computation in the following way: the  conformal tranformations
of the fields were calculated by hand and the computation of all
the correlation functions between these transformed fields was
done with the help of Mathematica. We have written a program to
compute the necessary  CFT correlation functions.

We denote (subscripts refer to the level)
\begin{eqnarray}
\widehat\Phi &=& t \widehat T + i r \widehat R+ i s \widehat S+
a \widehat A
+e  \widehat E +
f \widehat F+
 k \widehat K +  l \widehat L +m \widehat M+n  \widehat N +
 p \widehat P\,,\nonumber \\
S(\widehat\Phi ) &=& S_0 + S_{1/2}+S_1+S_{3/2} + S_2 +
S_{5/2}+S_3 + S_{7/2} + S_4\,. \nonumber
\end{eqnarray}
The factors of $i$ are included because of the reality condition
on the string field. We have obtained the following result:
\begin{eqnarray}
S_0 &=& \frac{t^2}{2}\,, \qquad S_{1/2}= \frac{9}{2}\ r\
t^2\,,\qquad
S_1 = -2\,r^2\,, \nonumber\\
S_{3/2} &=& 8\,r\,s\,t-10\,a\,t^2+2\,e\,t^2-20\,f\,t^2\,,\nonumber\\
S_2 &=& -\frac{1}{2} s^2\,,\nonumber\\
S_{5/2} &=&
\frac{16}{3\,\sqrt{3}}\,e\,{r^2}+\frac{10}{3}\,r\,s^2
-\frac{23}{3}\,k\,r\,t-\frac{25}{2}\,m\,r\,t+2\,n\,r\,t -p\,r\
t-\nonumber\\
& &-\, \frac{64}{3}\,a\,s\,t-\frac{80}{3}\,f\,s\,t\,,\nonumber
\end{eqnarray}
\begin{eqnarray}
S_3 &=&  -4\,a\,e-10\,f^2\,,\nonumber\\
S_{7/2}&=& -\frac{64}{3\,\sqrt{3}}\,a\,e\,r +\frac{32}{9\
\sqrt{3}}\,e^2\,r
-\frac{320}{9\,\sqrt{3}}\,e\,f\,r
+\frac{320}{9\,\sqrt{3}}\,f^2\,r-\nonumber\\
& & -\,8\,k\,r\,s +\frac{160}{27}\,l\,r\,s -\frac{100}{9}\,m\,r\,s
+\frac{80}{9}\,n\,r\,s +\frac{152}{27}\,p\,r\,s+\nonumber\\
& & +\,\frac{460}{27}\,a\,k\,t -\frac{28}{27}\,e\,k\,t
+\frac{280}{27}\,f\,k\,t +\frac{128}{27}\,a\,l\,t +\frac{64}{27}\
e\,l\,t +\frac{250}{9}\,a\,m\,t-\nonumber\\
& & -\,\frac{50}{9}\,e\,m\,t +\frac{220}{3}\,f\,m\,t
-\frac{104}{9}\,a\,n\,t +\frac{40}{9}\,e\,n\,t -\frac{880}{27}\
f\,n\,t+\nonumber\\
& & +\,\frac{20}{9}\,a\,p\,t -\frac{4}{9}\,e\,p\,t
-\frac{200}{27}\,f\,p\,t\,,\nonumber\\
S_4&=&-6\,{k^2}+6\,k\,l-3\,{l^2}-
\frac{45}{4}\,{m^2}+6\,{n^2}+\frac{3}{2}\,{p^2}\,.
\nonumber
\end{eqnarray}

Taking a look at the potential, we notice immediately that the
$Z_2$ twist symmetry of Berkovits' theory is absent in Witten's
theory. Indeed, if this symmetry were present the fields $r$ and
$t$ would carry twist charge $-1$ and $+1$, respectively and the
$r t^2$ term would be constrained to vanish. The origin of this
difference between the two theories can be traced to the
different structure of their lagrangians: in Berkovits' action,
all operators are inserted on the boundary of the disc, while in
Witten's action there is always an insertion of the
picture-changing operator in the origin. The proof of twist
invariance of the former theory~\cite{BSZ} breaks down for
Witten's theory since it makes use of transformations that do not
leave the origin invariant and hence move the picture-changing
operator away from the origin.

At level 1 and 2 the fields $r$ and $s$ can be integrated out
exactly to give the following effective potentials:
\begin{eqnarray}
V_0(t)&=&-\frac{{t^2}}{2}\,,\qquad
V_1(t)=-\frac{{t^2}}{2}-\frac{81\,{t^4}}{32}\,,\nonumber\\
V_2(t)&=&-\frac{{t^2}}{2}+\frac{81\,{t^4}}{2\,{{(4-64\,{t^2})}^2}}
-\frac{648\,{t^6}}{{{(4-64\,{t^2})}^2}}-\frac{81\,{t^4}}{4\,(4-64\
{t^2})}\,.\nonumber
\end{eqnarray}
We see that the inclusion of these higher modes does not alter
the fact that the potential has no minimum, in fact, its slope
becomes even steeper and a singularity is encountered at level 2,
see figure~\ref{f1}. The singularity in the tachyon potential
encountered here might be contrasted with the singularities in
the tachyon potential of open bosonic string theory found
in~\cite{bosonic}. In the case at hand, the potential diverges at
the singular point. Moreover, it doesn't have the interpretation
as a point where different branches of the effective  potential
come together since, at level 2, the equations for the fields
that are integrated out are linear.

Integrating out the fields numerically for the higher levels, one
finds that the more fields are included, the steeper the slope of
the potential becomes. This behaviour was anticipated in the
conclusions of~\cite{BSZ}.

\FIGURE{\epsfig{figure=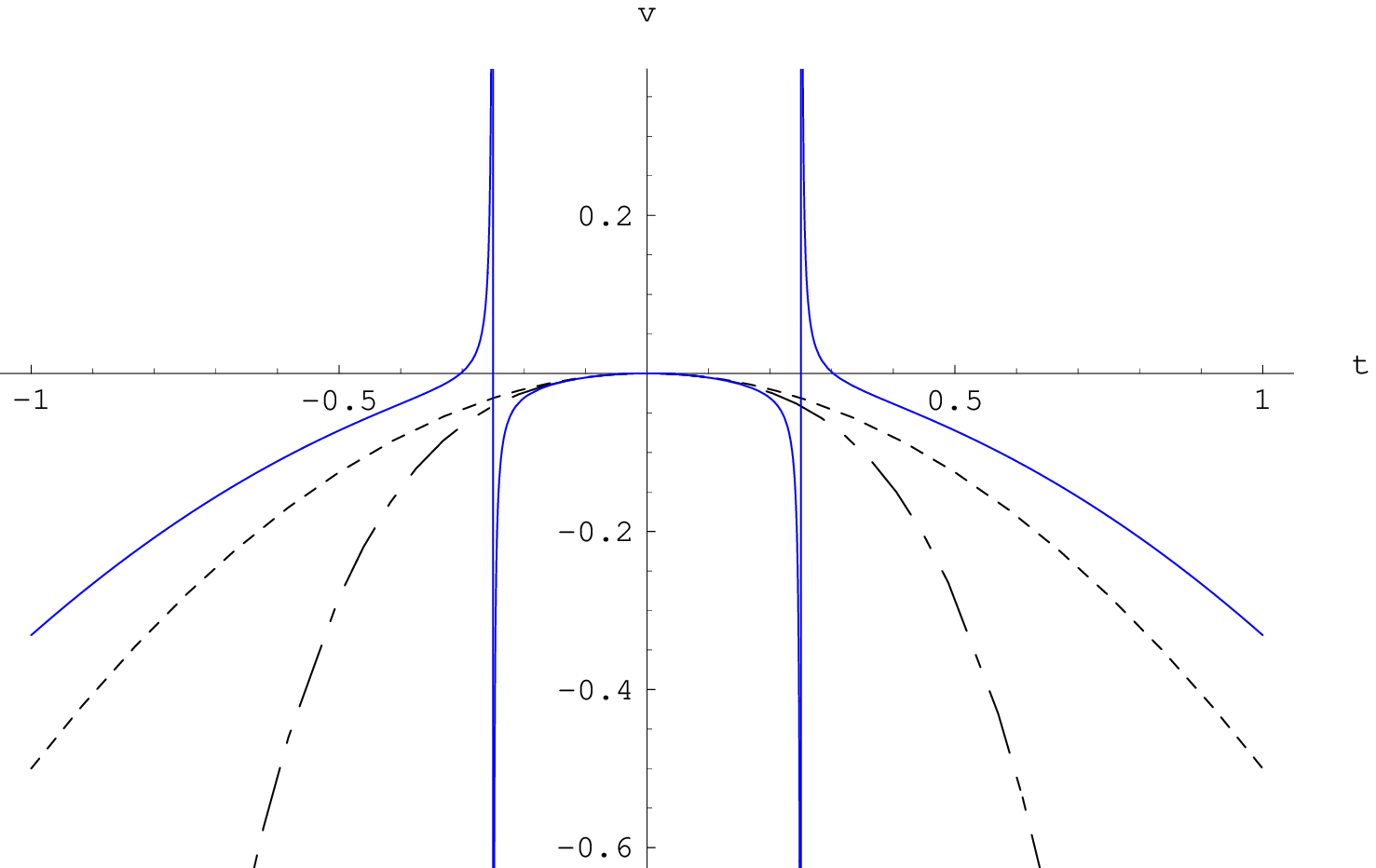,width=33em}%
\caption{The tachyon potential $V(t)$ at level~0 (dotted line),
level~1 (dashed line) and level~2 (full line).\label{f1}}}

\section{Conclusions}

In this letter we calculated the tachyon potential in the NS sector of
Witten's superstring field theory. We used a level truncation method
and kept terms up to level four in the potential. We found that the
potential does not exhibit the expected minimum, and that the
situation seems to become worse the more levels are included. This
result, in addition to other problems encountered within this
formulation, seems to point towards Berkovits' string field theory
proposal as a more viable candidate for the description of off-shell
superstring interactions.

\acknowledgments

This work was supported in part by the European Commission TMR
project \linebreak ERBFMRXCT96-0045. We would like to thank
B.~Zwiebach for discussions. P.J.D.S. is aspirant FWO-Vlaanderen.

\appendix

\section{The conformal transformations of the fields}
\label{transforms}

We now list the conformal transformations of the fields used in
the calculation of the tachyon potential. To shorten the notation
we denote $w = f(z)$.
\begin{eqnarray}
f \circ T(z) &=& (f'(z))^{-1/2} T(w)\,,\nonumber\\[2pt]
f \circ R(z) &=& R(w)\,,\nonumber\\[2pt]
f \circ S(z) &=& (f'(z))^{1/2} S(w) -
\frac{1}{2}\frac{f''(z)}{f'(z)}(f'(z))^{-1/2}c
e^{-\phi}(w)\,,\nonumber\\[2pt]
f \circ A(z) &=& f'(z) A(w) -
\frac{f''(z)}{f'(z)}c\partial c\,\partial\xi\,e^{-2 \phi}(w)
\,,\nonumber\\
f \circ E(z) &=& f'(z) E(w)\,,\nonumber \\
f \circ F(z) &=& f'(z) F(w)\,,\nonumber\\
f \circ K(z) &=& (f'(z))^{3/2} K(w) +
2 \frac{f''(z)}{f'(z)} (f'(z))^{1/2} c
\partial\left( e^{-\phi}\right)(w)+\nonumber\\
& &   + \left[ \frac{1}{2}\frac{f'''}{f'}-\frac{1}{4}
\left(\frac{f''}{f'}\right)^2 \right] (f'(z))^{-1/2}
c e^{-\phi}(w)\,,\nonumber\\
f \circ L(z) &=& (f'(z))^{3/2} L(w)+
\frac{f''(z)}{f'(z)}(f'(z))^{1/2}
c \partial\phi\,e^{-\phi}(w)+\nonumber\\
& & + \left[ \frac{3}{4}\left(\frac{f''}{f'}\right)^2-
\frac{2}{3}\frac{f'''}{f'}\right] (f'(z))^{-1/2}
c e^{-\phi}(w)\,,\nonumber\\
f \circ M(z) &=& (f'(z))^{3/2} M(w) +
\frac{15}{12} \left[\frac{f'''}{ f'} -
\frac{3}{2}\left( \frac{f''}{f'}\right)^2\right](f'(z))^{-1/2}
c e^{-\phi}(w)\,,\nonumber\\
f \circ N(z) &=& (f'(z))^{3/2} N(w) -  \frac{f''(z)}{f'(z)}(f'(z))^{1/2}
\partial c\, e^{-\phi}(w)+\nonumber\\
& & + \left[2 \left(\frac{f''}{f'}\right)^2-
\frac{f'''}{f'}\right] (f'(z))^{-1/2}
c\,e^{-\phi}(w)\,,\nonumber\\
f \circ P(z) &=& (f'(z))^{3/2} P(w)+\nonumber \\
& & + \left[ \frac{1}{4} \left(\frac{f''}{f'}\right)^2-
\frac{1}{6}\frac{f'''}{f'}\right] (f'(z))^{-1/2}  c e^{-\phi}(w)
\,.\nonumber
\end{eqnarray}

\section{Cyclicity property of string amplitudes}
\label{proofcyclicity}

The proof of the cyclicity is based on appendix A of~\cite{BSZ}.
First we give some preliminary remarks.

It is easy to see that all the fields in the GSO($+$) sector are
tensored with either $I$ or $\sigma_3$, and all the fields in the
GSO($-$) sector with either $\sigma_1$ or $i
\sigma_2$. The fields in the GSO($+$) sector have integer conformal
weight, and the fields in \pagebreak[3] GSO($-$) sector have
half-integer conformal weight. We compute

\begin{eqnarray}
\llll \hp_1 \cdots \hp_n \rrr
&=&\Tr \left\langle
\widehat{Y}
 f_1^n \circ \hp_1\widehat{X}  \cdots \widehat{X} f_{n-1}^n \circ
\hp_{n-1} \widehat{X} f_{n}^n
 \circ \hp_{n} \right\rangle \nonumber\\
&=&\Tr \left\langle
\widehat{Y}
 f_2^n \circ \hp_1\widehat{X}  \cdots \widehat{X} f_{n}^n \circ
\hp_{n-1} \widehat{X}\,R\circ f_{1}^n
 \circ \hp_{n} \right\rangle,\nonumber
\end{eqnarray}
where $R$ is the rotation over an angle of $2 \p$. Next we use
$R\circ f_{1}^n\circ \hp_{n}=\pm  f_{1}^n\circ \hp_{n}$, with a
plus sign if the field has integer weight, and a minus sign if
the field has half-integer weight. We also use the cyclicity of
the trace to move the field $\hp_n$ in front. Although we are
manipulating grassmann objects, we do not get an additional minus
sign, due to the fact that the total amplitude is grassmann odd.
\begin{eqnarray}
\llll \hp_1 \cdots \hp_n \rrr&=&\pm\Tr \left\langle
\widehat{Y}  f_{1}^n
\circ \hp_{n}
f_2^n \circ \hp_1\widehat{X}  \cdots \widehat{X} f_{n}^n \circ
\hp_{n-1} \widehat{X}\right\rangle\nonumber\\
&=&\pm\Tr \left\langle
\widehat{Y} \widehat{X} \circ f_{1}^n
\circ \hp_{n}
f_2^n \circ \hp_1\widehat{X}  \cdots \widehat{X} f_{n}^n \circ
\hp_{n-1} \right\rangle\nonumber\\
&=&\Tr \left\langle
\widehat{Y}   f_{1}^n
 \circ \hp_{n} \widehat{X}
 f_2^n \circ \hp_1\widehat{X}  \cdots \widehat{X} f_{n}^n \circ
\hp_{n-1} \right\rangle\nonumber\\
&=&\llll \hp_n \hp_1 \cdots \hp_{n-1} \rrr.
\end{eqnarray}
In the next to last line the picture changing operator commutes
with the GSO($+$) fields and anticommutes with the GSO($-$)
fields, cancelling the minus sign in front of the amplitude.

\end{document}